\documentclass[%
reprint,
amsmath,amssymb,
prb,
]{revtex4-1}

\usepackage{graphicx}
\usepackage{dcolumn}
\usepackage{bm}

\usepackage{hyperref} 
\hypersetup{
	colorlinks=true,
	linkcolor=blue,
	citecolor=red,
	filecolor=blue,      
	urlcolor=blue,
}
\usepackage{hypcap}

\usepackage{ulem} 
\usepackage[dvipsnames]{xcolor} 

\newcommand{\review}[1]{{\leavevmode\color{black}#1}}

\graphicspath{{Images/}}

\setcitestyle{numbers,square}

\begin{document}

\title{Effects of a non-causal electromagnetic response on the linear momentum transfer from a swift electron to a metallic nanoparticle}

\author{J. Castrej\'on-Figueroa}%
\email{jcastrejon@ciencias.unam.mx}
\author{J. \'A. Castellanos-Reyes}
\author{A. Reyes-Coronado}
\affiliation{Departamento de F\'isica, Facultad de Ciencias, Universidad Nacional Aut\'onoma de M\'exico, Ciudad Universitaria, Av. Universidad $\# 3000$, Mexico City, 04510, Mexico.}

\date{December 13, 2021}


\begin{abstract}

Electron beams in \review{Scanning} Transmission Electron Microscopes (STEMs) can be used as a tool to induce movement on nanoparticles.
Employing a classical-electrodynamics approach, it has been reported that the linear momentum transfer from a STEM-beam electron to a metallic spherical nanoparticle can be either repulsive or attractive towards the swift electron trajectory.
\review{This is in qualitative agreement with experimental observations.
The interaction time between a swift electron and a nanoparticle is typically on the order of attoseconds. 
Hence, the electromagnetic response of the nanoparticle at short times is of utmost importance.
However, it has been reported that the dielectric function employed in previous studies presented a non-causal pre-echo at the attosecond timescale, which might have led to incorrect unphysical results.}
Therefore, the validity of these linear momentum transfer results should be revisited.
\review{In this theoretical work, we study the non-causality effects on the linear momentum transferred from a swift electron to a metallic nanoparticle,} made of either aluminum or gold. 
Using an efficient numerical methodology, we found that non-causality, \review{as well as deficient numerical convergence, may lead} to incorrect repulsive linear momentum transfer results.
Contrary to what previous theoretical studies have \review{reported}, our results show that the linear momentum transfer from a swift electron to \review{spherical aluminum and gold nanoparticles, with radius 1\,nm, is always attractive.
Hence, a theoretical description of the experimentally observed repulsive interaction is pending.}

\end{abstract}

\maketitle

\section{Introduction}

The manipulation of micro- and nano-objects has been a field of interest since the second half of the last century given its potential applications for new technologies \citep{Morita,Romo-Herrera,Volpe,Dholakia,Ashkin,Wuite,Ali,Dholakia2}. 
In particular, it has been experimentally observed that the Scanning Transmission Electron Microscope (STEM) can be used as a tool to induce movement on nanoparticles (NPs) due to forces of repulsion or attraction toward the electron beam \citep{Oleshko2,Batson2,NanoLet,HZheng,Oleshko,TChen,YLiu,SGwo,Ticos} \review{and improved technologies in the reduction of the probe size} \citep{Haider2,Batson,Erni,krivanek1}.

The interaction between spherical NPs and aloof STEM electron beams has been theoretically addressed using a classical-electrodynamics approach by solving Maxwell’s equations in frequency space\textemdash called hereafter fully retarded wave solution \citep{GAbajo99,GAbajoRev,AbajoFJ}. In particular, this interaction has been studied through the \review{mechanical} linear momentum transferred by a single swift electron to a small metallic NP \citep{GAbajo,PRB2010,PRB2016,Ultramicroscopy,MgO_2019}, which involves the calculation of a closed-surface integral around the NP and an integral in the whole frequency space. These previous studies have shown that the linear momentum transferred to an aluminum or gold NP is predominantly attractive towards the swift electron trajectory but becomes repulsive at small impact parameters. This is in qualitative agreement with what has been experimentally observed \cite{Batson2,NanoLet,HZheng,Oleshko,TChen,YLiu,SGwo,Ticos}.

\review{The interaction between a swift electron and a NP typically occurs at the attosecond timescale \citep{PRB2016,PRB2021}.
Therefore, the electromagnetic response of the NP at this timescale is of utmost importance for the description of this interaction.
However, it has been reported that the dielectric function employed in previous linear momentum transfer calculations for gold NPs presented a non-causal pre-echo of tens of attoseconds \cite{causality}, which might have led to incorrect unphysical results. 
This non-causality is due to the interpolation and extrapolation of data, collected from different experiments (with different samples) and carried out for different frequency ranges, compiled by Palik \citep{Palik}.}
\review{Therefore, the validity of the previous linear momentum transfer results should be revisited.}
Additionally, the Newton-Cotes rules used in \review{previous} works \review{(see for example Refs. \citep{PRB2010,PRB2016,Ultramicroscopy,MgO_2019})} have a slow convergence rate (compared to a Gaussian quadrature) and do not provide error estimations without further calculations \cite{kronrod}, \review{leading\textemdash in certain situations\textemdash to incorrect physical conclusions}. 
\review{Hence, it is necessary to examine and improve the numerical methodology employed to compute the linear momentum transfer.} 
%
%

In this work, we \review{revise the non-causality effects on} the linear momentum transferred from a swift electron to a metallic NP, made of either aluminum or gold.
We use the fully retarded wave solution to Maxwell's equations, with the Gauss-Kronrod quadrature \cite{kronrod,kronrod2} and Double-Exponential formulas for numerical integration \cite{mori,mori2},
\review{presenting an efficient numerical methodology that provides results with accurate error estimates. }

%

In this work, we use SI units \review{unless} otherwise stated.


\section{Fully retarded wave solution approach} \label{Fully}

\begin{figure}
\centering
\includegraphics[width=0.45\textwidth]{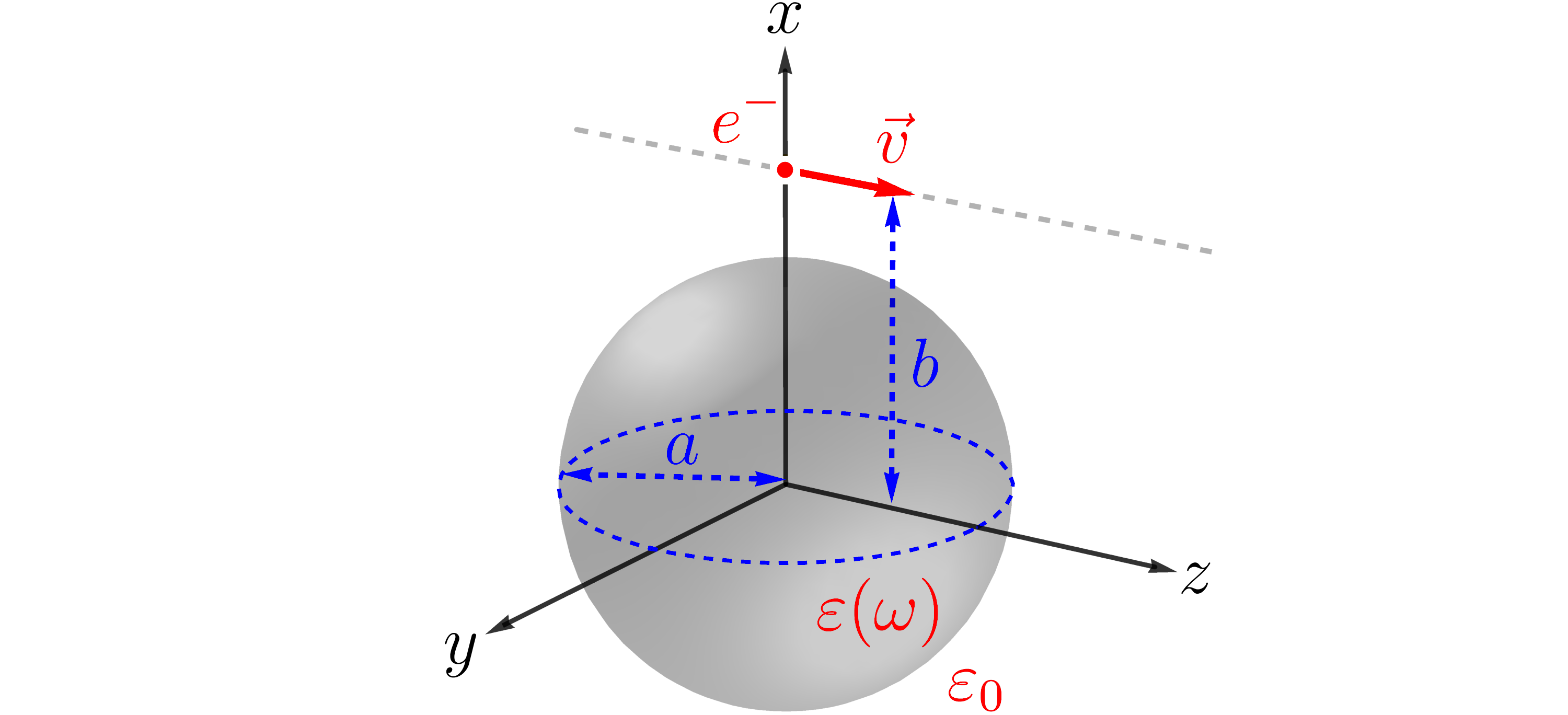}
\caption{Metallic nanoparticle (grey sphere) of radius $a$, \review{characterized with a frequency-dependent} dielectric function $\varepsilon (\omega)$, and embedded in vacuum, interacting with a swift electron (red dot) traveling in the $z$ direction with constant velocity $\vec{v}$ and impact parameter $b$.}
\label{Fig1}
\end{figure}

The electron beams produced in modern STEMs can reach an energy of up to 400\,keV \citep{GAbajoRev}. These beams consist of electrical currents on the order of tens of pA, equivalent to a train of swift electrons traveling in a straight trajectory with constant speed ($\sim0.83c$, with $c$ the speed of light), each one emitted approximately every $10^{-8}$\,s. Given that the typical lifetime of electronic excitations in metals is $\sim 10^{-14}$\,s \cite{Quijada}, we can safely assume that the NP interacts with a single swift electron at a time \cite{GAbajo99,GAbajoRev,AbajoFJ}. We consider the NP as an uncharged non-magnetic sphere embedded in vacuum, with radius $a$ and characterized with a \review{frequency-dependent} dielectric function $\varepsilon(\omega)$. We set the NP center as the origin of a Cartesian coordinate system, and we assume that the electron (a classical point particle with electric charge $-e$) travels with a constant velocity $\vec{v}$ along the $z$ direction, at a distance $b$ (impact parameter) from the origin, as shown in Fig. \ref{Fig1}. 
As it has been argued in Ref. \citep{AbajoFJ}, under these conditions, a quantum description of the incident electron and the NP is not necessary. Thus, in this work we employ a classical-electrodynamics approach to describe the interaction between the swift electron and the NP.

The fully retarded wave solution corresponds to the exact solution to Maxwell's equations in frequency space. Within this approach, the electromagnetic fields scattered by the spherical NP are obtained as a multipole expansion in spherical coordinates \citep{GAbajo99}:
\begin{align}
\vec{E}^{\,\text{scat}} (\vec{r};\omega) = \sum_{\ell = 1}^{\infty} \sum_{m=-\ell}^{m=\ell}
E_{\ell m}^r \hat{r} + E_{\ell m}^\theta \hat{\theta} + E_{\ell m}^\phi \hat{\phi}, \label{Escat} \\
\vec{H}^{\text{scat}} (\vec{r};\omega) = \sum_{\ell = 1}^{\infty} \sum_{m=-\ell}^{m=\ell}
H_{\ell m}^r \hat{r} + H_{\ell m}^\theta \hat{\theta} + H_{\ell m}^\phi \hat{\phi}. \label{Hscat} 
\end{align}
From these scattered fields, whose detailed expressions can be found in Ref. \citep{PRB2021}, together with the electromagnetic fields produced by the bare swift electron \review{(see Ref. \citep{MacielWave})}, one can obtain the total linear momentum transferred by the swift electron to the NP \review{by integrating the linear momentum conservation law in time, which can be written in the frequency space as} \cite{PRB2021}
\begin{equation} 
\label{momentum}
\Delta\vec{P}
=
\int_{0}^{\infty} \vec{\mathcal{P}}(\omega) \, d\omega ,
\end{equation}
with the spectral contribution to the linear momentum transfer, \review{$\vec{\mathcal{P}}(\omega)$}, given as
\begin{align} \label{spectral}
&\vec{\mathcal{P}}(\omega)\!=
\frac{1}{\pi} \!\oint_S \!\text{Re}\bigg\{\!
\varepsilon_0 \Big[\vec{E}(\vec{r};\omega)\vec{E}^{*}(\vec{r};\omega) \!-\! \frac{\tensor{I}}{2}\vec{E}(\vec{r};\omega) \! \cdot \!\vec{E}^{*}(\vec{r};\omega) \Big] \nonumber \\
& \,\, + \! \mu_0 \Big[\vec{H}(\vec{r};\omega)\vec{H}^{*}(\vec{r};\omega) \! - \! \frac{\tensor{I}}{2}\vec{H}(\vec{r};\omega)\! \cdot \!\vec{H}^{*}(\vec{r};\omega) \Big] \! \bigg\} \! \cdot \! d\vec{a},
\end{align}
%
where $S$ is a closed surface enclosing the NP that does not intersect the electron path. Here, $\text{Re}[z]$ denotes the real part of $z$, \review{$\tensor{I}$ is the unit dyadic}, $\varepsilon_0$ and $\mu_0$ are the electric permittivity and magnetic permeability of vacuum, respectively, and $\vec{E}\left(\vec{r};\omega\right)$ and $\vec{H}\left(\vec{r};\omega\right)$ are the total electromagnetic fields\textemdash the sum of those produced by the swift electron and those scattered by the NP. We refer the reader to Ref. \cite{PRB2021} for a detailed derivation of Eqs. (\ref{momentum}) and (\ref{spectral}).

\review{To compute the linear momentum transferred by the swift electron to the NP, it is customary to first calculate $\vec{\mathcal{P}}(\omega)$ from the closed-surface integral in Eq. (\ref{spectral}). Then, $\Delta\vec{P}$ is obtained through the frequency integral of $\vec{\mathcal{P}}(\omega)$ over $\omega\in(0,\infty)$ [see Eq. (\ref{momentum})].}

\review{It has been reported that the NP could be repelled or attracted towards the trajectory of the swift electron \citep{GAbajo,PRB2010,PRB2016,Ultramicroscopy,MgO_2019}. In particular, information about this attraction (or repulsion) is contained in the transverse ($x$ direction in our case, see Fig. \ref{Fig1}) component of the linear momentum transferred, $\Delta P_{\bot}$.  Thus, we focus our analysis and calculations on this component. 
Therefore, in Eqs. (\ref{momentum}) and (\ref{spectral}), we calculate both surface and frequency integrals numerically ensuring that at least the first three significant digits of $\Delta P_{\bot}$ are correct. The detailed numerical methodology, valid for NPs of any radius, is presented in Appendix \ref{AppendixA}. In particular, we consider up to 30 multipoles in Eqs. (1) and (2) ($\ell_\text{max}=30$) for all the calculations of $\Delta P_\bot$ in this work.}

In the next Sections, we calculate $\Delta P_\bot$ for a NP with $a=1$\,nm, made of either aluminum (Drude model) or gold (dielectric function taken from experimental data). It is noteworthy that these cases have been previously reported, showing that $\Delta P_\bot$ is negative\review{\textemdash repulsive\textemdash}at small impact \review{parameters \citep{GAbajo,PRB2010,PRB2016,Ultramicroscopy,MgO_2019}.} However, \review{as we show in Appendix \ref{Appendix B},} we found that numerical convergence for $\Delta P_\bot$ was not achieved \review{in those works, yielding incorrect negative $\Delta P_\bot$ values.}


\section{Linear momentum transferred by a swift electron to an aluminum nanoparticle} \label{aluminum}

\begin{figure*}
\includegraphics[width=1\textwidth]{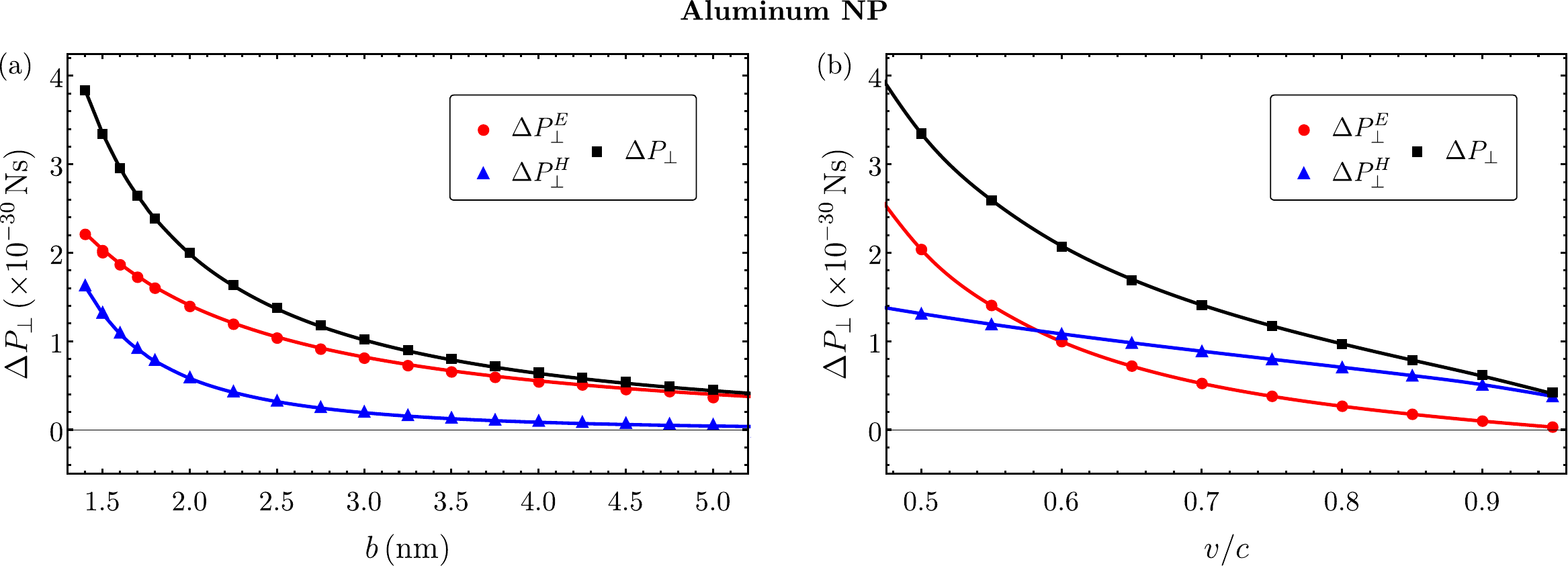}
\caption{Transverse linear momentum ($\Delta P_{\bot}$, black squares) transferred by a swift electron to an aluminum NP with $a=1$\,nm, (a) as a function of $b$ with $v=0.5c$, and (b) as a function of $v/c$ with $b=1.5$\,nm. The red circles and blue triangles represent the electric and magnetic contributions to $\Delta P_{\bot}$, respectively. The red, blue, and black lines are shown as a guide to the eye.}
\label{Fig. DPAl_perp}
\end{figure*}

In this Section, we calculate the transverse linear momentum, $\Delta P_{\bot}$, transferred from a swift electron to an aluminum NP with radius $a=1$\,nm.  \review{The dielectric function for aluminum was considered as the one given by the Drude model, with parameters $\hbar \omega_p = 15.1$\,eV and $\hbar \Gamma  = 0.15$\,eV, which are the same used in Ref. \cite{PRB2010}.
Although size corrections to the bulk dielectric function are important for metallic NPs with 1\,nm radius \citep{Kreibig}, we found they yield a negligible contribution to the linear momentum transfer \citep{PRB2021}. Thus, in this work we always use bulk dielectric functions to characterize the NP.}

In Fig. \ref{Fig. DPAl_perp}(a) we show $\Delta P_{\bot}$ as a function of \review{the impact parameter} $b$ with \review{an electron speed} $v=0.5c$, and in Fig. \ref{Fig. DPAl_perp}(b) $\Delta P_{\bot}$ as a function of $v/c$ with $b=1.5$\,nm.
We separate $\Delta P_{\bot}$ (black squares) in its electric ($\Delta P_{\bot}^E$, red circles), and magnetic ($\Delta P_{\bot}^H$, blue triangles) contributions [first and second terms, inside square brackets, on Eq. (\ref{spectral}), respectively]. We highlight three characteristics of Fig. \ref{Fig. DPAl_perp}: (i) $\Delta P_{\bot}$ and its electric and magnetic contributions are always positive, \review{meaning an effective} attractive interaction between the electron and the NP; (ii) $\Delta P_{\bot}$ decreases for larger values of both $b$ [see Fig. \ref{Fig. DPAl_perp}(a)] and $v$ [see Fig. \ref{Fig. DPAl_perp}(b)]; and (iii) at $b$ = 1.5 nm and $v/c \approx 0.58$, the electric and magnetic contributions to $\Delta P_{\bot}$ cross each other [see Fig. \ref{Fig. DPAl_perp}(b)]. Thus, at $v/c \lesssim 0.58$ the electric contribution to $\Delta P_{\bot}$ dominates over the magnetic one, even if the impact parameter increases [see Fig. \ref{Fig. DPAl_perp}(a)]. Conversely, at $v/c \gtrsim 0.58$ the magnetic contribution becomes larger than the electric one, and at $v=0.95c$ most of the linear momentum is transferred by the magnetic contribution.

Our findings show that $\Delta P_{\bot}$ for an aluminum NP with radius $a$ = 1\,nm is always positive, \review{meaning that} the NP is attracted towards the swift electron trajectory. This is in contradiction with previous \review{theoretical} results reporting $\Delta P_{\bot}$ as negative at $b=1.5$\,nm and $v=0.5c$ \cite{PRB2010}.
\review{To investigate the origin of the discrepancies between our results and the ones reported in Ref. \cite{PRB2010}, we compared the numerical methodologies employed in the calculations, showing that it is possible to obtain incorrect repulsive results for $\Delta P_{\bot}$ if the number of sampling points in the frequency space is low enough so that numerical convergence is not achieved (see Appendix \ref{Appendix B} for the details).}

\review{The causality of the dielectric function is of utmost importance for the calculation of $\Delta\vec{P}$ \cite{causality}, since it is necessary to integrate over all frequencies [see Eq. (\ref{momentum})].} The dielectric function used to characterize the electromagnetic response of the aluminum NP\textemdash the Drude model\textemdash is causal, that is, it fulfills Kramers-Kronig relations \cite{causality}. However, it has been suggested that a non-causality in the dielectric function may also lead to incorrect repulsive results \citep{causality}. Thus, \review{in the following section}, we study the unphysical effects on the linear momentum transfer caused by a non-causal response of the NP. \review{In particular, we focus on the case of a small nanoparticle of radius 1\,nm. Therefore, we study the effects on the linear momentum transfer computations caused by a non-causality in the quasistatic polarizability. }

\begin{figure*}
\includegraphics[width=0.98\textwidth]{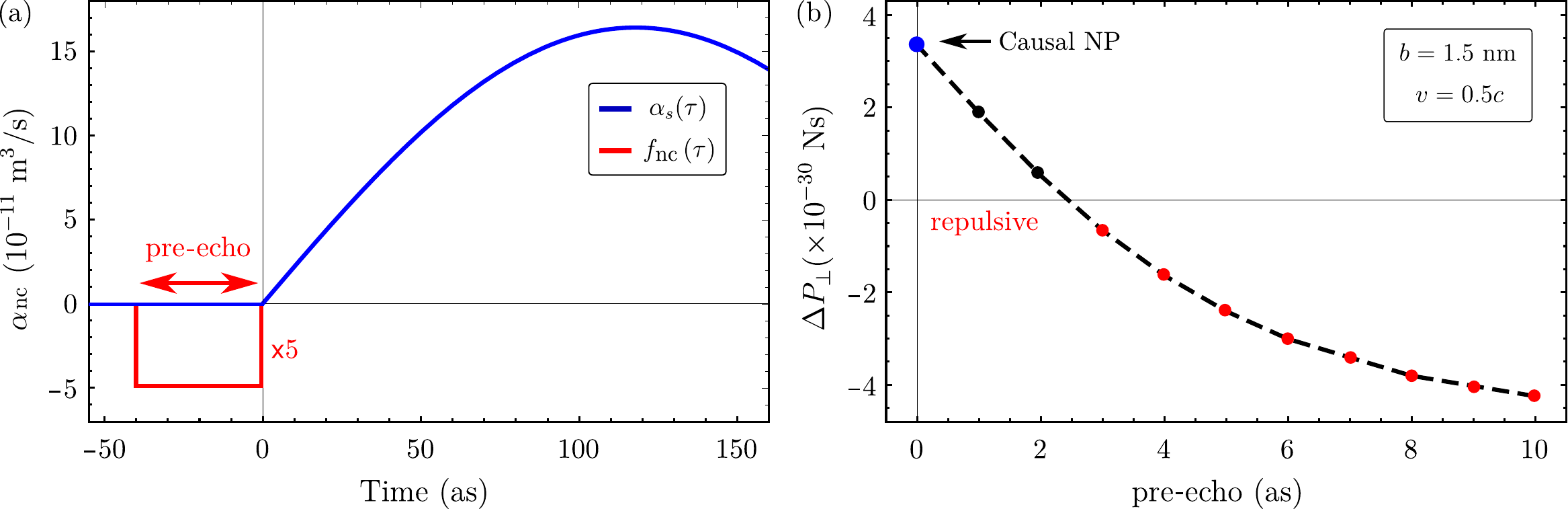}
\caption{(a) Non-causal polarizability $\alpha_{\text{nc}}(\tau)$ as a function of time, given in Eq. (\ref{alpha_nc}), for a NP with radius $a=1$\,nm. The red line represents the non-causal contribution $f_{\text{nc}}(\tau)$, see Eq. (\ref{fnc_t}), added to the quasistatic polarizability $\alpha_s(\tau)$ (blue line) given in Eq. (\ref{alpha_t}). (b) Transverse linear momentum transferred, $\Delta P_\bot$, to the same NP as in (a), as a function of the pre-echo, with $b=1.5$\,nm and $v=0.5c$, \review{considering up to $\ell_\text{max}=30$ in Eqs. (\ref{Escat}) and (\ref{Hscat})}. The blue dot corresponds to $\Delta P_\bot$ transferred to an aluminum NP characterized with a causal dielectric function (with no pre-echo), while the red dots correspond to a repulsive interaction due to the non-causal response of the NP. The black dash line is a guide to the eye.}
\label{Fig. DP_NC}
\end{figure*}

\section{Unphysical repulsive interaction caused by a non-causal electromagnetic response of the NP}

As has been argued in Ref. \cite{PRB2021}, the electromagnetic response of small spherical NPs is satisfactorily characterized by the quasistatic polarizability: 
\begin{equation}\label{alpha_w0}
\alpha_s (\omega)= 4 \pi a^3 \dfrac{\varepsilon(\omega) - \varepsilon_0}{\varepsilon(\omega) + 2\varepsilon_0},
\end{equation}
where $\varepsilon(\omega)$ is the \review{frequency-dependent} dielectric function of the NP and $a$ its radius. From Eq. (\ref{alpha_w0}), it is possible to express $\varepsilon(\omega)$ in terms of $\alpha_s(\omega)$ as
\begin{equation}\label{eps_w}
\dfrac{\varepsilon(\omega)}{\varepsilon_0} =  \dfrac{2\alpha_s(\omega)  +  4 \pi a^3}{ 4 \pi a^3 - \alpha_s(\omega)}.
\end{equation}

For an aluminum NP with dielectric function given by the Drude model (with the parameters chosen in Section III), the quasistatic polarizability is given by
\begin{equation}\label{alpha_w}
\alpha_s (\omega)= 4 \pi a^3 \dfrac{\omega_s^2}{\omega_s^2 - \omega^2-i\omega\Gamma},
\end{equation}
with $\omega_s=\omega_p/\sqrt{3}$. By means of a frequency-to-time Fourier transform, we calculate the quasistatic polarizability as a function of time:
\begin{align}
\alpha_s(\tau) &= \dfrac{1}{2\pi}\int_{-\infty}^{\infty} \alpha_s(\omega) e^{-i\omega\tau} d\omega \nonumber \\
&=  4 \pi a^3 \Theta(\tau) \dfrac{\omega_s^2}{\Omega_s} e^{- \tau \Gamma  /2} \sin(\Omega_s \tau), \label{alpha_t}
\end{align}
where $\Theta(\tau)$ is the Heaviside step function, and $\Omega_s = \omega_s \sqrt{1-(\Gamma/2\omega_s)^2}$. Since $\alpha_s(\tau<0) = 0$ due to $\Theta(\tau)$, the quasistatic polarizability given in Eq. (\ref{alpha_t}) is causal. It is possible to add a controlled artificial non-causality $f_{\text{nc}}$ to $\alpha_s$, defining a non-causal quasistatic polarizability $\alpha_{\text{nc}}$ as
\begin{align}\label{alpha_nc}
\alpha_{\text{nc}} = \alpha_s + f_{\text{nc}}.
\end{align}
For simplicity, we choose $f_{\text{nc}}(\tau)$ as a boxcar function:
\begin{align}\label{fnc_t}
f_{\text{nc}}(\tau) = \left\{ \begin{array}{cc} 
                -\text{A} & \hspace{5mm} \tau \in (-\text{T},0\,) \\
                0 & \hspace{5mm}  \text{other case} 
                \end{array} , \right.
\end{align}
with length T and height A. By means of a time-to-frequency Fourier transform we obtain
\begin{equation}\label{fnc_w}
f_{\text{nc}}(\omega) = - \text{AT} \, \text{sinc} \Bigg( \dfrac{\omega \text{T}}{2} \Bigg) e^{-  i  \omega \text{T}/2},
\end{equation}
with $\text{sinc}(z)=\sin(z)/z$. Notice from Eq. (\ref{fnc_w}) that T has units of time and AT has units of volume.

Interestingly, in Ref. \citep{GAbajo}, it was developed an expression for the transverse linear momentum transferred in the limit $b \gg a$, given in cgs units as
\begin{equation}\label{dp_small}
\Delta p_x = \dfrac{e^2}{b^4 v} \Bigg( 5.55165 + \dfrac{1.85055}{\gamma} \Bigg) \text{Re}[\alpha(\omega=0)],
\end{equation}
with $e$ the fundamental electric charge and $\gamma=[1-(v/c)^2]^{-1/2}$. From Eqs. (\ref{alpha_w}), (\ref{alpha_nc}), and (\ref{fnc_w}), one can notice that
\begin{equation}\label{alpha_0}
\alpha_{nc} (\omega=0) = 4 \pi a^3 - \text{AT}.
\end{equation}
Hence, if the volume AT of the non-causality is larger than $4\pi a^3$, then $\Delta p_x$ is negative [see Eqs. (\ref{dp_small}) and (\ref{alpha_0})], meaning that the NP will be repelled from the electron trajectory.

To corroborate if a non-causal polarizability yields a repulsive interaction within the fully retarded wave solution approach, from Eq. (\ref{eps_w}) we define the non-causal dielectric function as
\begin{equation}\label{eps_nc}
\dfrac{\varepsilon_{\text{nc}}(\omega)}{\varepsilon_0} =   \dfrac{2\alpha_{\text{nc}}(\omega)  +  4 \pi a^3}{ 4 \pi a^3 - \alpha_{\text{nc}}(\omega)},
\end{equation}
and we further calculate $\Delta P_{\bot}$ to a NP characterized with a dielectric function given by Eq. (\ref{eps_nc}).

In Fig. \ref{Fig. DP_NC}(a) we show the polarizability given in Eq. (\ref{alpha_nc}) as function of time, for a NP characterized with the dielectric function given by Eq. (\ref{eps_nc}), \review{with the Drude (aluminum) parameters,} and with radius $a=1$\,nm. We separate $\alpha_{\text{nc}}(\tau)$ in its causal [$\alpha_s(\tau)$ given by Eq. (\ref{alpha_t}), blue line] and non-causal [$f_{\text{nc}}(\tau)$ given by Eq. (\ref{fnc_t}), red rectangle] contributions. We refer to the length T of the red rectangle as the pre-echo and we choose $\text{A}=10^{-11}\,\text{m}^3/\text{s}$ in Eq. (\ref{fnc_t}) (same order of $\alpha_s$).
In Fig. \ref{Fig. DP_NC}(b) we show $\Delta P_{\bot}$ transferred to the NP as a function of the pre-echo, choosing $b=1.5$\,nm and $v=0.5c$, \review{considering up to $\ell_\text{max}=30$ in Eqs. (\ref{Escat}) and (\ref{Hscat}) to ensure numerical convergence (see Appendix A)}.

Previous studies have reported that most of the linear momentum is transferred from the electron to the NP in a time scale on the order of tens of attoseconds \citep{PRB2016,PRB2021}, where the main interaction between the swift electron and the NP occurs. A remarkable effect can be observed when the non-causal pre-echo is on the order of the interaction time between the NP and the swift electron: the transverse linear momentum transferred is negative \review{[red dots in Fig. \ref{Fig. DP_NC}(b)], meaning the NP is repelled from the swift electron trajectory.}
However, \review{these negative values of $\Delta P_\bot$ are} the result of considering a non-causal dielectric function for the NP, and thus \review{they are} an artifact.

\review{There have been studies reporting negative values of $\Delta P_{\bot}$ for other materials \citep{GAbajo,PRB2010,PRB2016,Ultramicroscopy,MgO_2019}}. In particular, negative values of $\Delta P_{\bot}$ for gold nanospheres \review{were reported} in Refs. \citep{PRB2010,Ultramicroscopy}. However, \review{it has recently been discovered that the dielectric function used in those works is not causal} \cite{causality}. Therefore, it is not clear that the reported negative values of $\Delta P_{\bot}$ for gold NPs are physical. In the next Section we study the effects  on $\Delta P_{\bot}$ due to the \review{non-causality} of the gold NP response.

\section{Linear momentum transferred from a swift electron to a gold nanoparticle}

\review{In this Section,} we consider a gold NP and \review{study} two cases for its dielectric function taken from experimental data: (i) a dielectric function from different experiments compiled by Palik \cite{Palik}, and (ii) a fitted dielectric function from experimental data reported by Werner \textit{et al}. \cite{Werner}.
We obtained $\Delta P_{\bot}$ with a precision of at least three significant digits (see Appendix \ref{AppendixA} for further details).

Since the dielectric function taken from experimental data compiled by Palik is only known in a limited frequency range, it is necessary to perform extrapolations (as well as interpolations) of the data, which does not necessarily yield a causal dielectric function \cite{causality}. Conversely, Werner \textit{et al}. fitted their experimental data with a superposition of Lorentz oscillators, resulting in an analytic causal dielectric function \cite{causality}.
To analyze the causality of both Palik and Werner \textit{et al}. dielectric functions, we calculate the quasistatic polarizability as a function of time through a numerical frequency-to-time Fourier transform.
%

\begin{figure}
\includegraphics[width=0.48\textwidth]{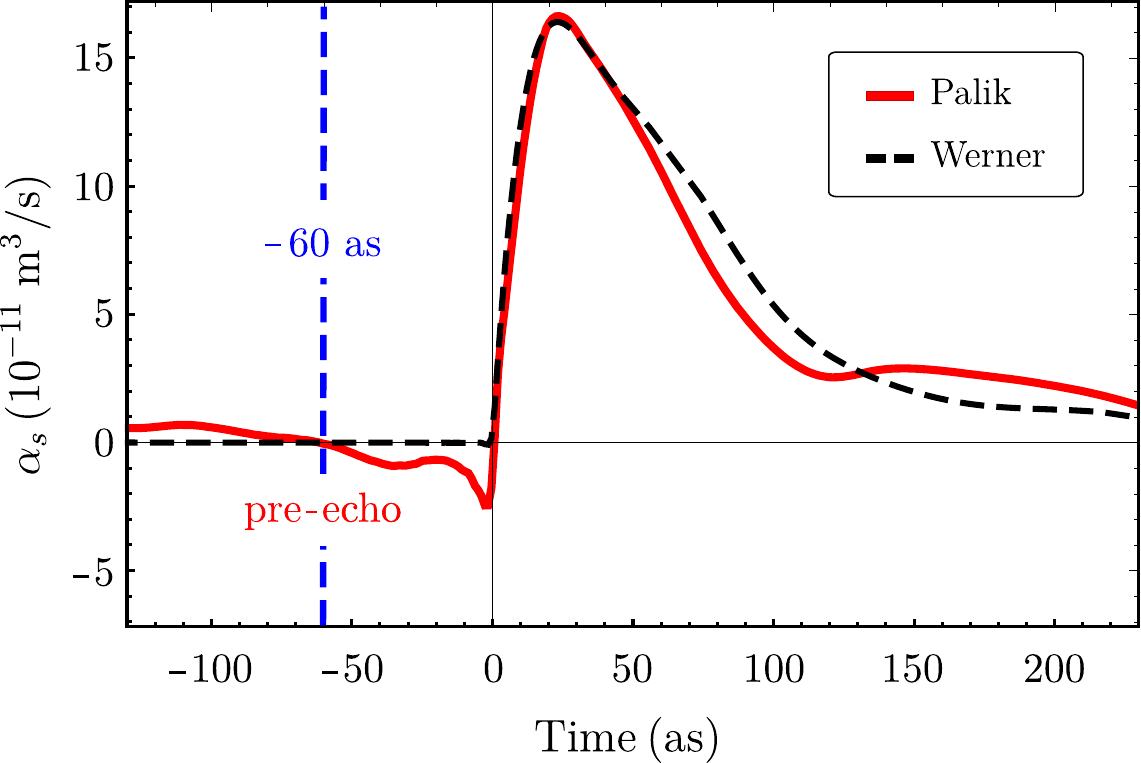}
\caption{Quasistatic polarizability as a function of time for a gold NP, with a dielectric function taken from experimental data compiled by Palik \citep{Palik} (red line), and with a dielectric function taken from Werner \textit{et al}. (black dash line) \cite{Werner}. The NP characterized by the dielectric function obtained from Palik data has a non-causal pre-echo for $\tau < 0$. The vertical blue dash line indicates $\tau=-60$\,as.}
\label{Fig. alpha_au}
\end{figure}

\begin{figure*}
\centering
\includegraphics[width=0.98\textwidth]{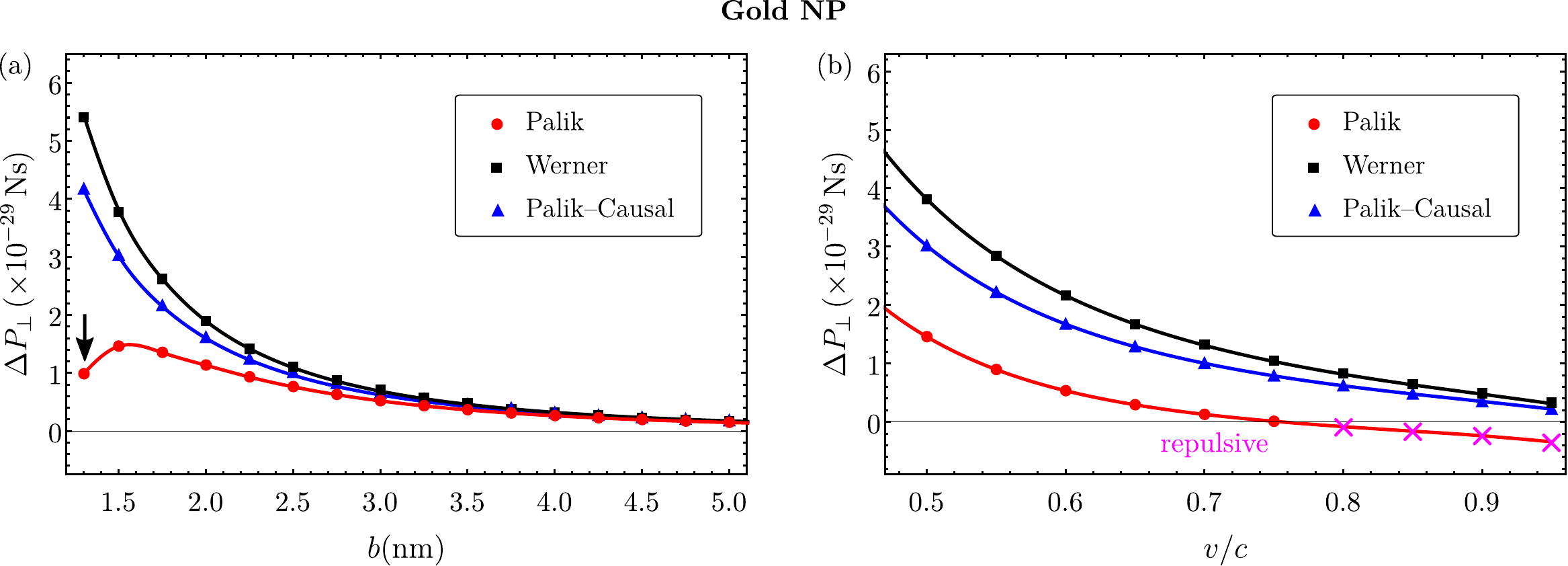}
\caption{Transverse linear momentum, $\Delta P_{\bot}$, transferred by a swift electron to a gold NP with $a=1$\,nm, (a) as a function of $b$ and with $v=0.5c$, and (b) as a function of $v/c$ and with $b=1.5$\,nm, \review{using $\ell_\text{max}=30$}. The red circles correspond to the results obtained considering a dielectric function taken from the experimental data compiled by Palik \cite{Palik}, the black squares correspond to a dielectric function taken from Werner \textit{et al}. \cite{Werner}, and the blue triangles correspond to the Palik dielectric function after forcing causality [Palik-Causal, Eq. (\ref{eps_causal})]. The vertical black arrow in (a) indicates $b=1.3$\,nm. In (b), the magenta crosses indicate where $\Delta P_{\bot} < 0$. The red, blue, and black lines are shown as a guide to the eye.}
\label{Fig. DP_Au}
\end{figure*}

We show in Fig. \ref{Fig. alpha_au} the time-dependent quasistatic polarizability for a gold NP calculated from the Palik data (red line), and from the Werner \textit{et al}. analytic function (black dash line). Since the dielectric function from Palik (coming from different experiments) is obtained through an interpolation of the data, the resulting time-dependent polarizability has a non-zero imaginary part and a pre-echo. We found that this pre-echo is present regardless of the interpolation method we employed. For simplicity, in the results presented in this work, we employed first-order polynomials to interpolate \review{the Palik data}.
In contrast, the time-dependent polarizability from Werner \textit{et al}. is a real function and is zero for $\tau<0$. This means that the dielectric function from Werner \textit{et al}. is causal, whereas the dielectric function from Palik data is not. The pre-echo in the Palik data (red line) shown in Fig. \ref{Fig. alpha_au} extends beyond -60 attoseconds (vertical blue dash line), which is on the same order of magnitude that the interaction time between the swift electron and the NP. As we show next, this pre-echo leads to incorrect repulsive results for the linear momentum transferred to the gold NP.

The non-causal quasistatic polarizability obtained from the data compiled by Palik $\alpha_s^{\text{P}}$ (red line in Fig. \ref{Fig. alpha_au}), can be converted into a causal polarizability by taking only its real part (in the time domain) and eliminating the pre-echo:
\begin{equation}\label{alfa_causal}
\hat{\alpha}_s^{\text{P}}(\omega) = \int_{-\infty}^{\infty} \Theta(\tau)\, \text{Re} \big[  \alpha_s^{\text{P}}(\tau)  \big] \, e^{i\omega\tau} \, d\tau,
\end{equation}
from which, using Eq. (\ref{eps_w}), one can obtain a causal dielectric function:
\begin{equation}\label{eps_causal}
\dfrac{\hat{\varepsilon}_{\text{P}}(\omega)}{\varepsilon_0}= \dfrac{2\hat{\alpha}_s^{\text{P}}(\omega)  +  4 \pi a^3}{ 4 \pi a^3 - \hat{\alpha}_s^{\text{P}}(\omega)}.
\end{equation}
From now on, we will refer to this causal dielectric function $\hat{\varepsilon}_{\text{P}}(\omega)$, obtained from Eq. (\ref{eps_causal}) and the experimental data for gold compiled by Palik, as Palik-Causal.

In Fig. \ref{Fig. DP_Au}(a) we show the transverse linear momentum transferred $\Delta P_{\bot}$ to a gold NP as a function of $b$ with $v=0.5c$, and in Fig. \ref{Fig. DP_Au}(b) as a function of $v/c$ with $b=1.5$ nm. Using the methodology described in \review{Appendix \ref{AppendixA}}, we calculate $\Delta P_{\bot}$ \review{(with $\ell_\text{max}=30$)} considering the three dielectric functions used to characterize the electromagnetic response of the gold NP: Palik (red circles), Werner (black squares), and Palik-Causal (blue triangles).
One can see in Fig. \ref{Fig. DP_Au}(a) that $\Delta P_{\bot}$ is positive for the three dielectric functions considered (meaning an attractive interaction between the NP and the electron). However, for the red circles (Palik), at $b=1.3$\,nm [indicated with a black arrow in \ref{Fig. DP_Au}(a)], it can be seen that
\begin{equation}
\Delta P_{\bot} (\small{b=1.3 \, \text{nm}}) < \Delta P_{\bot} (\small{b=1.5 \, \text{nm}}),
\end{equation}
showing a maximum in $\Delta P_{\bot}$, which is not observed for the causal dielectric functions (Werner and Palik-Causal). The value of $\Delta P_{\bot}$ at $b=1.3$ nm was previously reported as negative, indicating a repulsive interaction between the NP and the electron \cite{PRB2010,Ultramicroscopy}.

In Fig. \ref{Fig. DP_Au}(b) one can see that $\Delta P_{\bot}$ decreases for larger values of $v$. For Palik data (red circles), $\Delta P_{\bot}$ is positive up to $v \approx 0.75c$, after which it becomes negative (repulsive interaction, indicated with magenta crosses). This transition from attractive to repulsive linear momentum transferred disappears for Werner (black squares) and  Palik-Causal (blue triangles) dielectric functions.

Therefore, we conclude that non-causality  \review{leads} to incorrect repulsive linear momentum transfer results \review{for the gold NP considered. Moreover, in general, it can be concluded that non-causality and deficient numerical convergence may lead to incorrect repulsive values of the linear momentum transfer.}


\section{Conclusions}

We studied the linear momentum transferred from a swift electron to a nanoparticle with radius $a=1$\,nm, made of either aluminum or gold.
Using the fully retarded wave solution to Maxwell's equations, we presented an efficient numerical methodology that ensures that at least the first three significant digits of the transverse linear momentum transferred are correct.
We found that the transverse linear momentum transferred from a swift electron to a NP with radius of 1\,nm, made of \review{either} aluminum \review{or gold, is} always positive (meaning that the NP will be \review{effectively} attracted towards the swift electron trajectory).

We analyzed the effects on the linear momentum transfer caused by a non-causal dielectric function (characterizing the NP) with an attoseconds pre-echo. 
\review{By controlling the magnitude of a non-causal pre-echo in the NP polarizability, 
we found that a non-causality} on the order of the interaction time (tens of attoseconds in our case) may lead to an unphysical repulsive interaction between the swift electron and the NP.

We calculated the linear momentum transferred from an electron to a gold NP using two dielectric functions: (i) one obtained from the data compiled by Palik \cite{Palik}, which we showed to be non-causal, and (ii) another, that turns \review{out} to be causal, fitted by Werner \textit{et al}. from their experimental data \cite{Werner}. For the non-causal case, we found a transition from attractive to repulsive interaction at $v \approx 0.75c$, with $b=1.5$ nm. This repulsive behavior is not present for a gold NP characterized by the causal dielectric function. Interestingly, we showed that if the  \review{non-causal} dielectric function is forced to be causal, the repulsive results disappear. Hence, we conclude that a non-causal dielectric function \review{leads} to incorrect repulsive results \review{for the gold NP}, as could be the case of Refs. \cite{PRB2010,Ultramicroscopy}.
The repulsive linear momentum transfer reported in previous theoretical works may have been caused by a combination of \review{using} a non-causal dielectric function and a deficient numerical convergence in the frequency integration. 
In fact, we have found that both causality and numerical convergence are essential to obtain physically sound results.

\review{The previously reported theoretical explanation of the experimentally observed repulsive interaction between STEM-electron beams and nanoparticles relied on the computed negative values of the linear momentum transfer. However, we have shown that the negative values for gold and aluminum NPs are unphysical and incorrect. Therefore, a theoretical description of the experimentally observed repulsion is pending.}

\begin{acknowledgments}
This work was supported by UNAM-PAPIIT project DGAPA IN114919. J. C-F. and J. \'A. C-R. are doctoral students from Programa de Doctorado en Ciencias Físicas, Universidad Nacional Autónoma de México (UNAM), and received scholarship 477516 and 481497, respectively, from Consejo Nacional de Ciencia y Tecnología (CONACyT), Mexico. We thank Rub\'en G. Barrera, Philip E. Batson, Andrea Konečná and Javier Aizpurua for valuable discussions and suggestions.
\end{acknowledgments}


\appendix

\section{Revisited numerical methodology} \label{AppendixA}

To compute the linear momentum transferred by the swift electron to the NP, we first calculate $\vec{\mathcal{P}}(\omega)$ from the closed-surface integral in Eq. (\ref{spectral}). Then, we obtain $\Delta\vec{P}$ through the frequency integral of $\vec{\mathcal{P}}(\omega)$ over $\omega\in(0,\infty)$ [see Eq. (\ref{momentum})]. \review{Both surface and frequency integrals are obtained numerically.}  

\review{As mentioned in the main text, the numerical methods employed in previous works might have led to incorrect results. For this reason, we revise the numerical methods used to calculate $\Delta P_{\bot}$, proposing a more precise methodology and ensuring that at least the first three significant digits of $\Delta P_{\bot}$ are correct.}

\subsection{Closed surface integral}

As mentioned in Section \ref{Fully}, the spectral contribution to the linear momentum transfer is given by \review{
\begin{align} \label{spectralA1}
\vec{\mathcal{P}}(\omega)=&
\frac{1}{\pi}\oint_S  \tensor{\mathcal{T}}(\vec{r};\omega)  \cdot d\vec{a},
\end{align}
with
\begin{align}
&\tensor{\mathcal{T}}(\vec{r};\omega) = \text{Re}\Big[ \varepsilon_0\vec{E}(\vec{r};\omega)\vec{E}^{*}(\vec{r};\omega) - \frac{\varepsilon_0}{2}\tensor{I}\vec{E}(\vec{r};\omega)\cdot\vec{E}^{*}(\vec{r};\omega) \nonumber \\
& +  \mu_0\vec{H}(\vec{r};\omega)\vec{H}^{*}(\vec{r};\omega) -\frac{\mu_0}{2}\tensor{I}\vec{H}(\vec{r};\omega)\cdot\vec{H}^{*}(\vec{r};\omega) \Big].
\end{align}
}
We choose the surface $S$ in Eq. (\ref{spectralA1}) as a spherical shell, concentric to the NP, with radius $R=a\,+\,0.05$\,nm, and we considered the spherical coordinate system ($r$,$\theta$,$\phi$), determined by the spherical-to-Cartesian transformation: $x=r \sin \theta \cos\phi$, $y=r \sin \theta \sin\phi$, $z=r\cos\theta$, with $(x,y,z)$ the Cartesian coordinate system shown in Fig. \ref{Fig1}. 
Equation (\ref{spectralA1}) can be written as
\begin{align} \label{spectral2}
\vec{\mathcal{P}}(\omega)&=
\frac{R^2}{\pi} \int_0^{2\pi} \!\! \int_0^{\pi} \tensor{\mathcal{T}}(R,\theta,\phi;\omega)  \cdot \hat{r} \sin \theta \, d\theta \, d\phi.
\end{align}
\review{As mentioned before, previous studies obtained $\vec{\mathcal{P}}(\omega)$ numerically using a composite Simpson $3/8$ rule, a Newton-Cotes (NC) rule of third-degree (see Refs. \cite{PRB2010,PRB2016,Ultramicroscopy,MgO_2019})}. However, a Gaussian quadrature has a faster converge rate than a NC rule with respect to the number of evaluations $N$ of the integrand \cite{kronrod}.

The Gauss-Kronrod quadrature (GKQ) is constructed by adding $N+1$ points to a $N$-point Gauss-Legendre quadrature (GLQ), resulting in a total of $2N+1$ nodes and weights \citep{kronrod,kronrod2}. The GKQ provides an error estimate by comparing the result of the integral with the one obtained using the nested $N$-point GLQ.
We define $g_j$ and $\theta_j$ as the weights and nodes for the integration of $\vec{\mathcal{P}}(\omega)$ in $\theta$, and $g_k$ and $\phi_k$ as the weights and nodes for the integration in $\phi$. 
Notice that the region of integration in Eq. (\ref{spectral2}) is the rectangle $[0,2\pi)\times[0,\pi]$.

The transverse component of the spectral contribution to the linear momentum transfer is given by
\begin{align} \label{Nspectral}
\mathcal{P}_\bot (\omega) =& \frac{R^2}{\pi} \sum_{k=1}^{{N_\phi}} \sum_{j=1}^{{N_\theta}} g_j g_k \mathcal{T}_\bot (R,\theta_j,\phi_k;\omega) \sin \theta_j \, \pm \epsilon_{\mathcal{P}_\bot}(\omega),
\end{align}
with \review{$\epsilon_{\mathcal{P}_\bot}$ the absolute numerical error, and} $\mathcal{T}_\bot = \hat{x} \cdot \tensor{\mathcal{T}} \cdot \hat{r}$, $N_\theta = 2n_\theta +1$ and $N_\phi = 2n_\phi +1$, where $n_\theta$ and $n_\phi$ correspond to the orders of the nested Gauss-Legendre quadratures.

To quantitatively compare the two methods of numerical integration: (i) the composite Simpson $3/8$ rule, and (ii) the Gauss-Kronrod quadrature (GKQ), we calculate the transverse component of the spectral contribution to the linear momentum transfer, $\mathcal{P}_{\bot} (\omega)$, for the aluminum NP with $a=1$ nm, \review{considering the Drude model with the parameters used in Section \ref{aluminum}}.
We show in Table \ref{Fig. IntS}, the relative error of $\mathcal{P}_{\bot}(\omega)$ as a function of the number of evaluations $N$ on the surface \review{$S$}.
In Table \ref{Fig. IntS}, $N=2n^2_{_{NC}}$ is the number of evaluations on the surface $S$ for the NC rule, with $n_{_{NC}}$ the number of sampling points for $\theta$ and $2n_{_{NC}}$ the number of sampling points for $\phi$. For the GKQ, $N \gtrsim  N_\theta \times N_\phi$ with $ N_\phi \approx N_\theta +10$. 
\review{It can be seen} that the GKQ has a smaller relative error ($\delta \mathcal{P}_{\bot}^{GKQ}$) than the NC rule ($\delta \mathcal{P}_{\bot}^{NC}$), indicating a faster convergence rate. 
Hence, in \review{all} our calculations, we employ the GKQ to calculate $\mathcal{P}_{\bot}$ with an error estimate $\epsilon_{\mathcal{P}_\bot}$.
\begin{table}[h!]
\includegraphics[width=0.23\textwidth]{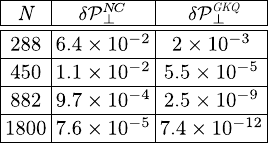}
\caption{Relative error of the closed-surface integral shown in Eq. (\ref{Nspectral}), as a function of the number of evaluations $N$ on the surface $S$, obtained using a Newton-Cotes (NC) rule of third degree ($\delta \mathcal{P}_{\bot}^{NC}$), and a Gauss-Kronrod quadrature ($\delta \mathcal{P}_{\bot}^{GKQ}$), for an aluminum \review{(Drude model)} NP with $a = 1$\,nm, and a swift electron with $b=1.5$\,nm and $v=0.5c$. We chose the dipole resonance frequency $\hbar \omega = 8.7$\,eV (arbitrarily) to illustrate this particular example.}
\label{Fig. IntS}
\end{table}
%


\subsection{Frequency integral}

The GKQ is useful to obtain the integral of functions on bounded intervals. To use the GKQ for the frequency integration given in Eq. (\ref{momentum}) it is convenient to define a cutting frequency $\omega_c$, so that the integral is calculated on $\left(0,\omega_c\right)$ instead of $\left(0,\infty\right)$. By defining $g_i$ and $\omega_i$ as a collection of weights and nodes of the GKQ for the frequency interval $(0,\omega_c)$, the linear momentum transferred from the swift electron to the NP can be written as
\begin{equation}\label{DP_GQK}
\Delta P_{\bot} = \sum_i g_i \mathcal{P}_{\bot}(\omega_i) \,  \pm  \, \epsilon_{P_{\bot}},
\end{equation}
where $\epsilon_{P_{\bot}}$ is the absolute numerical error.

\review{To estimate $\epsilon_{P_{\bot}}$ in Eq. (\ref{DP_GQK}), it is necessary to consider all the sources of numerical error, in particular, the numerical error of the spectral contribution to the linear momentum [see Eq. (\ref{Nspectral})].}
The absolute numerical error for the closed-surface integral $\epsilon_{\mathcal{P}_\bot}$ in Eq. (\ref{Nspectral}) is a function of the frequency $\omega$, and it is given by the difference between the results obtained with GKQs and  GLQs \cite{kronrod}. The contribution of $\epsilon_{\mathcal{P}_\bot}$ to the linear momentum transfer is given by the integral of $\epsilon_{\mathcal{P}_\bot}$ over the frequency interval $(0,\omega_c)$. However, we only need to estimate an upper bound to such error:
\begin{equation}\label{errS}
\omega_c \, \text{Max} \big[ \epsilon_{\mathcal{P}_\bot} (\omega_i) \big] \gtrsim \int_0^{\, \omega_c} \epsilon_{\mathcal{P} _\bot } (\omega) \, d\omega,
\end{equation}
where the index $i$ runs over the sum for the frequency integral [shown in Eq. (\ref{DP_GQK})]. Hence, the error due to the closed-surface numerical integration can be expressed as $\omega_c \, \text{Max}[ \epsilon_{\mathcal{P}_\bot}]$.

\review{Then,} the absolute numerical error $\epsilon_{P_{\bot}}$ is given by
\begin{equation}\label{error}
\epsilon_{P_{\bot}} =  \epsilon_{_{GKQ}} + \omega_c \, \text{Max}[ \epsilon_{\mathcal{P}_\bot} (\omega_i)] + 
\tilde{\epsilon}_{cut},
\end{equation}
with $\epsilon_{_{GKQ}}$ the estimation of the error provided by the GKQ for the frequency integration on the interval $(0,\omega_c)$, $\omega_c \, \text{Max}[ \epsilon_{\mathcal{P}_\bot} (\omega_i)]$ the closed-surface integration error, and $\tilde\epsilon_{cut}$ an upper bound to the error by frequency truncation:
\begin{equation}\label{Etail}
\tilde\epsilon_{cut} \gtrsim \epsilon_{cut} = \int_{\omega_c}^{\infty} \mathcal{P}_{\bot}(\omega)\,  d\omega.
\end{equation}

\begin{figure}
\includegraphics[width=.48\textwidth]{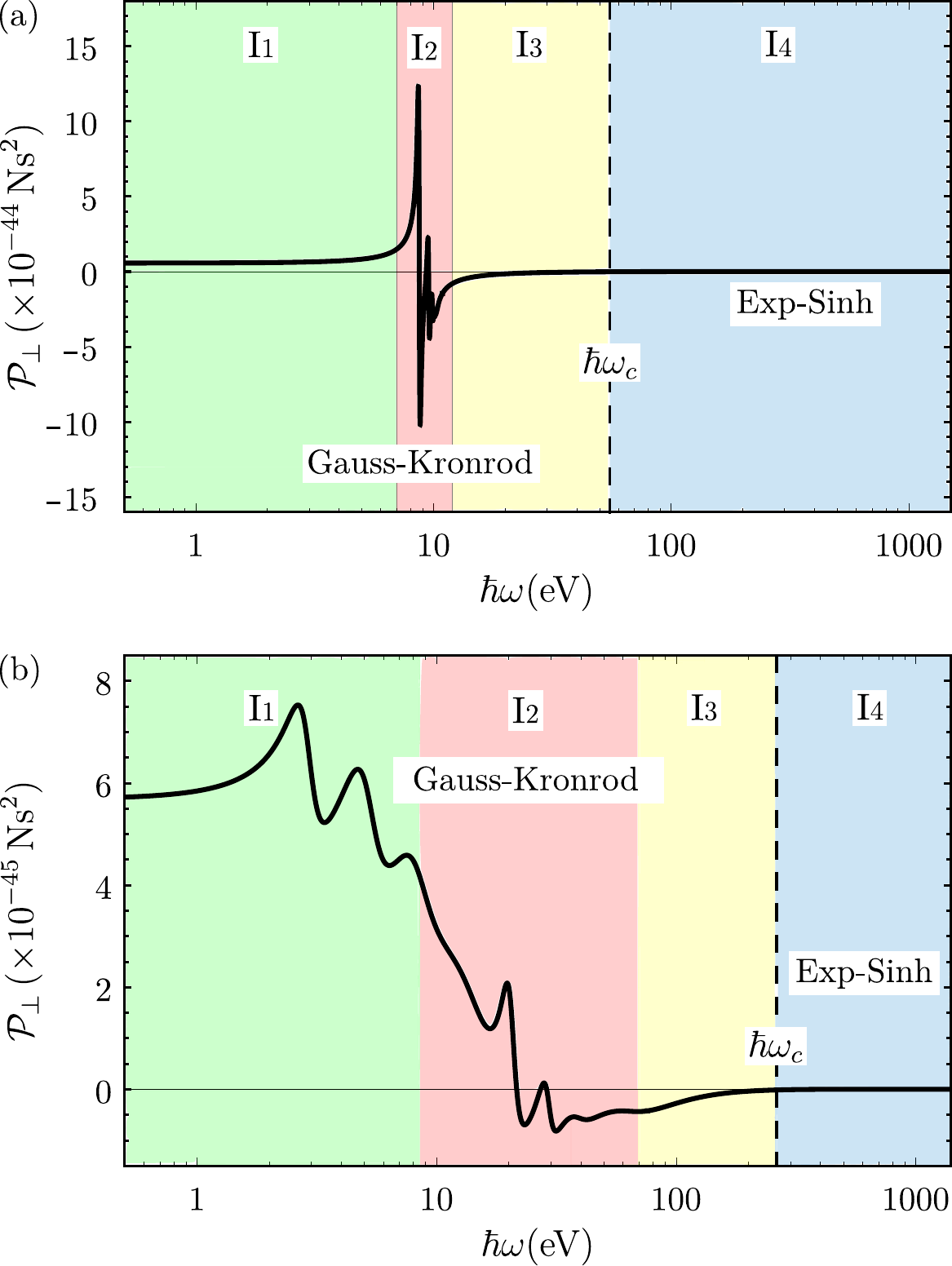}
\caption{Typical spectral contribution to the transverse linear momentum transfer, $\mathcal{P}_{\bot}(\omega)$, (a) for an aluminum \review{(Drude model)} NP, and (b) for a gold NP \cite{Werner}. Four relevant frequency regions are shown in colors: I$_1$ in green, I$_2$ in red, I$_3$ in yellow, and I$_4$ in blue. The blue region is used to estimate $\omega_c$ and $\tilde{\epsilon}_{cut}$ using the Exp-Sinh quadrature, while the green, red, and yellow regions, are integrated using the GKQ.}
\label{Fig. dPdwAl}
\end{figure}

Next, we discuss the criteria for choosing $\omega_c$ and the collection of weights and nodes ($g_i$ and $\omega_i$) in Eq. (\ref{DP_GQK}).
\review{In Fig. \ref{Fig. dPdwAl}(a) we show a typical spectral contribution to the transverse linear momentum, $\mathcal{P}_{\bot}(\omega)$, for the aluminum NP (Drude model), and in Fig. \ref{Fig. dPdwAl}(b) we show a typical $\mathcal{P}_{\bot}(\omega)$ for gold NP (Werner).} We have separated the entire frequency range into four regions labeled from $\text{I}_1$ to $\text{I}_4$, and highlighted with colors: green, red, yellow, and blue, respectively. 
In the red region of Fig. \ref{Fig. dPdwAl}(a), $\text{I}_2$, $\mathcal{P}_{\bot}(\omega)$ shows a rich peak structure in contrast to the other regions (green, yellow, and blue), in which $\mathcal{P}_{\bot}(\omega)$ varies smoothly. \review{Analogously, in the red region of Fig. \ref{Fig. dPdwAl}(b), $\mathcal{P}_{\bot}(\omega)$ changes from positive to negative.
For this reason, a greater number of sampling points is needed in the red regions than in the remaining ones.} 
As mentioned before, the GKQ is used to compute $\Delta P_{\bot}$ in $\left(0,\omega_c\right)$ instead of $\left(0,\infty\right)$. To calculate the error by frequency truncation, $\epsilon_{cut}$ [see Eq. (\ref{Etail})], at a given cutoff frequency $\omega_c$, we employ the Exp-Sinh quadrature (Double-Exponential method) \cite{mori,mori2} implemented in the BOOST libraries \cite{boost} to obtain the integral of $\mathcal{P}_{\bot}(\omega)$ in $\left(\omega_c,\infty\right)$ (blue region in Fig. \ref{Fig. dPdwAl}). It is worth mentioning that the Exp-Sinh quadrature has convergence limitations for rapidly changing integrands, like the \review{ones} shown in the red \review{regions} of Fig. \ref{Fig. dPdwAl}, hence we only applied it for the blue regions.

\begin{figure}
\includegraphics[width=.48\textwidth]{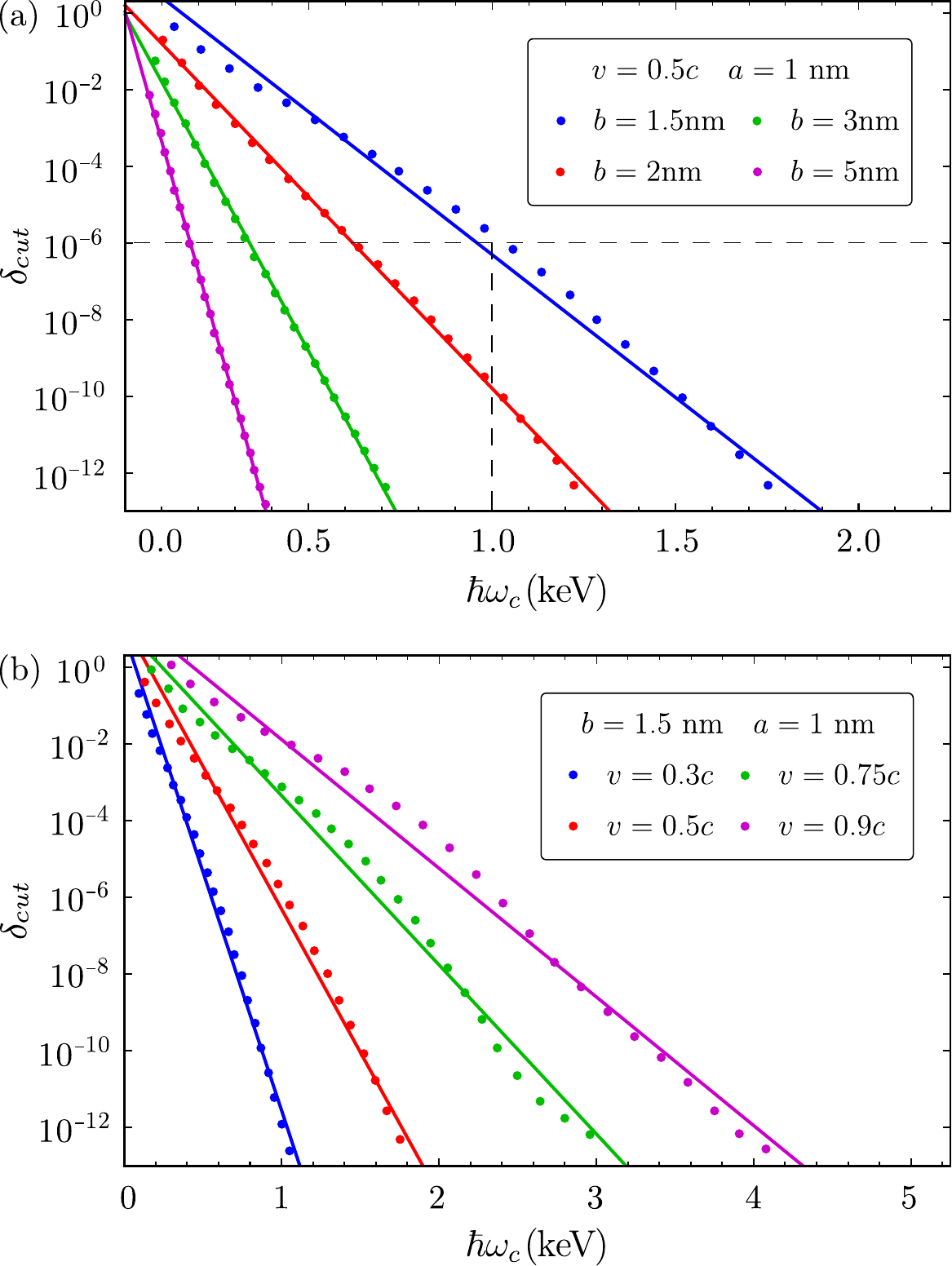}
\caption{Relative error by frequency truncation of the linear momentum transferred, $\delta_{cut} = \epsilon_{cut} / \Delta P_{\bot}$, as a function of $\hbar \omega_c$. (a) $\delta_{cut}$ for different impact parameters $b$, with $v=0.5c$, and (b) $\delta_{cut}$ for different speeds of the electron $v$, with $b=1.5$\,nm. It is shown a fitted line to the calculated dots as a guide to the eye.}
\label{Fig. IntW}
\end{figure}

In Fig. \ref{Fig. IntW} we show the relative error by frequency truncation, $\delta_{cut} = \epsilon_{cut} / \Delta P_{\bot}$, as a function of $\hbar \omega_c$, \review{for the aluminum (Drude model) NP with the parameters indicated in the inset.}
One can notice that the inclination of the fitted line decreases when $b$ decreases [Fig. \ref{Fig. IntW}(a)] and $v$ increases [Fig. \ref{Fig. IntW}(b)]. 
The \review{nformation shown} in Fig. \ref{Fig. IntW} can be used as a guide to replicate, with a given error tolerance, \review{the results presented in the main text.}
For example, to obtain $\Delta P_{\bot}$ with $\tilde\epsilon_{cut} \sim 10^{-6} \Delta P_{\bot}$ for an electron with $v=0.5c$ and $b=1.5$\,nm, it is necessary to integrate $\mathcal{P}_{\bot}$ up to $\hbar \omega_c \approx 1$\,keV [shown with black dash lines in Fig. \ref{Fig. IntW}(a)]. \review{The gold NP case is completely analogous.}

After choosing $\omega_c$ so that $\delta_{cut}$ has the desired tolerance, we select the weights and nodes, $g_i$ and $\omega_i$, in Eq. (\ref{DP_GQK}) and calculate $\Delta P_{\bot}$ using a GKQ in the remaining regions $\text{I}_1$ to $\text{I}_3$ (see Fig. \ref{Fig. dPdwAl}). 
\review{For our results, we selected $N_\theta = 83$ and $N_\phi=91$ for the GKQ in the closed surface integral, achieving a relative numerical error of $\delta_{\mathcal{P}_\bot} \sim 10^{-11}$.
For the aluminum NP we choose GKQ orders of 51, 201, and 151 (equivalent to 103, 403, and 303 sampling points) for the green, red, and yellow regions shown in Fig. \ref{Fig. dPdwAl}(a), respectively.
For the gold NP we choose GKQ orders of 101, 181, and 101 (equivalent to 203, 363, and 203 sampling points) for the green, red, and yellow regions shown in Fig. \ref{Fig. dPdwAl}(b), respectively.
We also considered $\ell_{\text{max}} = 30$ in Eqs. (\ref{Escat}) and (\ref{Hscat}), and we choose $\omega_c$ [accordingly to the Fig. \ref{Fig. IntW}] so that $\delta_{cut} < 10^{-6}$ in each case. 
By this means, in all our results we ensure a numerical precision $\epsilon_{P_ \bot}  < 10^{-4} \Delta P_{\bot}$ [see Eq. (\ref{DP_GQK})], which means that at least the first three significant digits of $\Delta P_{\bot}$ are correct. 
In our calculations, we observed that the main contribution to the error in Eq. (\ref{error}) is $\epsilon_{GK}$, which means that the number of sampling points on $\omega\in(0,\omega_c)$ determines if the numerical convergence is achieved, assuming $\hbar \omega_c$ is on the order of kiloelectronvolts and a sufficiently large number of sampling points for the surface $S$ is considered. }

\begin{figure}
\includegraphics[width=0.48\textwidth]{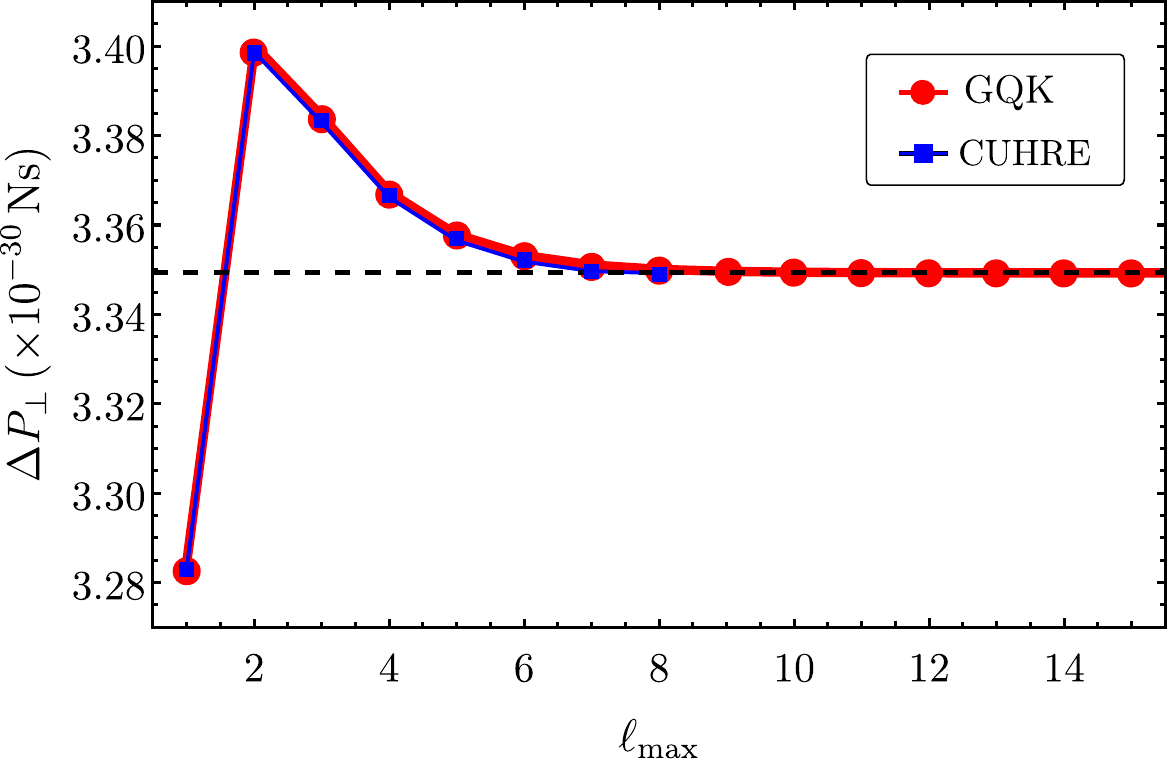}
\caption{Transverse linear momentum, $\Delta P_\bot$, transferred from a swift electron with $v=0.5c$ and $b=1.5$\,nm to an aluminum NP with $a=1$\,nm, as a function of the maximum number of multipoles, $\ell_\text{max}$, considered in the scattered fields [Eqs. (\ref{Escat}) and (\ref{Hscat})]. The blue squares correspond to $\Delta P_\bot$ obtained with the CUHRE algorithm \cite{cuba}, while the red circles to the one obtained with the GKQ. The horizontal black dash line indicates the value of $\Delta P_\bot$ with $\ell_\text{max}=30$ obtained with GKQ. The blue and red lines are shown as a guide to the eye. 
}
\label{Fig. dp_lmax}
\end{figure}

To corroborate the accuracy of the numerical method employed, we quantitatively validate our results with the CUHRE algorithm \cite{cuba}, which provides an automatic integrator reporting accurate error estimates. However, due to the high computational demand of this method, we only applied it to a few points. In Fig. \ref{Fig. dp_lmax} we show $\Delta P_\bot$ as a function of the maximum number of multipoles, $\ell_\text{max}$, for an aluminum NP with $a=1$\,nm, $b=1.5$\,nm and $v=0.5c$, using both methods: the GKQ (red circles) and CUHRE (blue squares). It can be seen that when $\ell_\text{max}$ increases, both methods converge to the same value, \review{indicated with a horizontal black dash line. At $\ell_\text{max}=8$, the difference between the two methods and the black dash line is in the fourth significant digit. Therefore, in this case, choosing $\ell_\text{max} \geq 8$ ensures the desired accuracy of $\Delta P_\bot$. In general, the specific value of $\ell_\text{max}$ depends on the radius of the NP and the impact parameter.}

\section{Incorrect repulsive results caused by deficient numerical convergence} \label{Appendix B}

\review{To investigate the origin of the discrepancies between our results and the ones reported in Ref. \cite{PRB2010}, we now compare the numerical methodologies used in the calculations of the frequency integral [Eq. (\ref{momentum})]. As discussed before, we divided the spectral contribution to the linear momentum transfer into four regions. The region $\text{I}_4$ was used to determine the cutoff frequency $\omega_c$, and the other three regions, $\text{I}_1$ to $\text{I}_3$, were integrated with the GKQ, using different quadrature orders on each region to achieve a numerical convergence of at least three significant digits in $\Delta P_{\bot}$. The main contribution of negative (repulsive) linear momentum transfer comes from the region $\text{I}_3$ (see yellow region in Fig. \ref{Fig. dPdwAl}(a); notice the logarithmic scale). In Ref. \cite{PRB2010}, the frequency integral was calculated using a composite Simpson $3/8$ rule. Hence, to understand why a repulsive interaction was previously obtained, we integrate the region $\text{I}_3$ using a composite Simpson $3/8$ rule and a fixed high order GKQ in the regions $\text{I}_1$ and $\text{I}_2$.
In Fig. \ref{Fig. dp_n3} we show the transverse linear momentum transferred, $\Delta P_{\bot}$, as a function of the number of sampling points $N_3$ in the region $\text{I}_3$. As can be seen, when the number of sampling points is below $N_3$=21, $\Delta P_{\bot}$ is negative, incorrectly indicating that the NP will be repelled from the swift electron trajectory. However, when the number of sampling points increases, $\Delta P_{\bot}$ converges asymptotically to the result obtained using the GKQ.
In Fig. \ref{Fig. dp_n3}, we indicate with a red dot the repulsive value previously reported in Ref. \citep{PRB2010}, which we infer to be incorrect due to a lack of convergence in the frequency integral, caused by a low number of sampling points.

\begin{figure}
\includegraphics[width=0.48\textwidth]{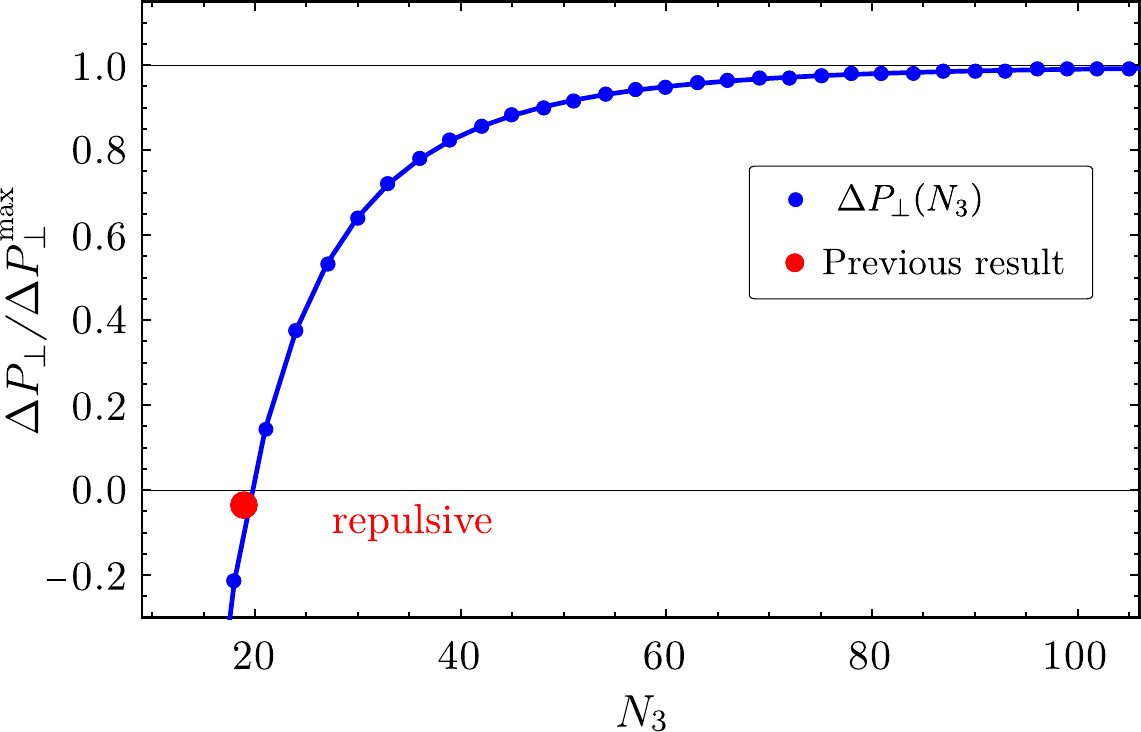}
\caption{Transverse linear momentum, $\Delta P_\bot$, transferred from a swift electron with $v=0.5c$ and $b=1.5$\,nm to an aluminum NP with $a=1$\,nm, as a function of the number of sampling points $N_3$ in the region $\text{I}_3$, and obtained using a composite Simpson $3/8$ rule. $\Delta P^{\text{max}}_{\bot}$ is the result obtained with the GKQ. The blue line is a guide to the eye. The previously repulsive result for the linear momentum transferred reported in Ref. \cite{PRB2010} is shown as a red dot.}
\label{Fig. dp_n3}
\end{figure}

Interestingly, it is possible to obtain incorrect repulsive results for the gold NP, as in the case of the aluminum NP, if the number of sampling points in the frequency space is low enough so that numerical convergence is not achieved.}

\end{document}